\documentclass[12pt]{iopart}
\usepackage{epsfig}
\usepackage{rotating}
\usepackage{portland}
\usepackage{lscape}
\usepackage{sw20lart}
\usepackage[dvips]{tcigraph}

\input{tcilatex}
\begin{document}

\title{Time scales in {LISA}.}
\author{Sophie Pireaux\thanks{}}

\begin{abstract}
The LISA mission is a space interferometer aiming at the detection of
gravitational waves in the [$10^{-4}$,$10^{-1}$] Hz frequency band. In order
to reach the gravitational wave detection level, a Time Delay Interferometry
(TDI) method must be applied to get rid of (most of) the laser frequency
noise and optical bench noise. This TDI analysis is carried out in terms of
the coordinate time corresponding to the Barycentric Coordinate Reference
System (BCRS), TCB, whereas the data at each of the three LISA stations is
recorded in terms of each station proper time. We provide here the required
proper time versus BCRS time transformation.\newline
We show that the difference in rate of station proper time versus TCB is of
the order of $5\cdot 10^{-8}$. The difference between station proper times
and TCB exhibits an oscillatory trend with a maximum amplitude of about $%
10^{-3}$ s.
\end{abstract}


\rotdriver{dvips}

\address{UMR 6162, ARTEMIS\newline
Observatoire de la C\^{o}te d'Azur,\newline
avenue de Copernic,\newline
06130 GRASSE,\newline
FRANCE\newline
Tel: ++33(0)4 93 40 53 70\newline
Fax: ++33(0)4 93 40 53 33}

\ead{sophie.pireaux@obs-azur.fr}

\submitto{\CQG}

\pagebreak

\section{Introduction}

\qquad The LISA mission \cite{LISA2000} is a space interferometer aiming at
the detection of gravitational waves (GW) in the $[\sim 10^{-4},\sim
10^{-1}] $ Hz frequency band. Gravitational waves crossing the LISA quasi
equilateral triangular constellation are detected through the induced change
in the station inter-distances. The latter also depend on time due to the
gravitational field of the Sun \cite{Chau2005}, mostly, and planets; what we
call a ``geometry (G) effect''.\newline
In order to reach the gravitational wave detection level, a Time Delay
Interferometry (TDI) method (see \cite{Tin2005} for a review and references therein) 
must be applied to get rid of (most of) the laser frequency (LF) noise and
optical bench (OB) noise. In other words, TDI is needed to bring those LF
plus OB (physically indistinguishable) noises down to the level of the other
noises : quantum (Q), fiber (F), residual proof mass motion (PM) noises... The
TDI method consists in combining numerically data fluxes at the stations
(rather than combining the laser beams physically) with an appropriate
delay. Hence, the so-called TDI observables are closed loop combinations of
the different laser links with appropriate delays (combination of
photon-flight time $t_{ij}$ between two stations $i\rightarrow j$ which
correspond to station inter-distances) that cancel (almost all) the laser
frequency noise and optical bench noise.\newline
The TDI analysis is carried out in terms of the coordinate time $t$
corresponding to the Barycentric Coordinate Reference System (BCRS), the
so-called TCB, whereas the data at each of the three LISA stations, $k=1,2,3$%
, is archived in terms of the station proper time $\tau _{k}$.\newline
We modeled the orbitography of the three stations ($\overrightarrow{x_{i}}%
,\ \overrightarrow{v_{i}}\equiv d\overrightarrow{x_{i}}/dt$ with $t=$TCB and 
$i=1,2,3$) \emph{classically} \cite{Dhu2005} in the gravitational field of
the sole Sun, and the laser links, that is the photon-flight time $t_{ij}$, 
\emph{relativistically} \cite{Chau2005} as a function of the position and
velocities of stations $i$ and $j$ at emission time.\newline
We here provide the corresponding analytical proper time versus TCB
transformation required to apply the TDI analysis. We show that the
difference between station proper time and TCB reaches about $0.5$ s over a
one-year mission and exhibits an oscillatory trend with a maximal amplitude
of $0.0014$ s. The difference in rate of station proper time versus TCB is
of the order of $1.5\cdot 10^{-8}$s.\\
We then show how this proper time versus TCB transformation fits in a general 
relativistic TDI analysis.


\section{Proper time versus coordinate time transformation}



\subsection{Classical orbitography}


\qquad We assume a classical orbitography for the three LISA\ spacecraft $%
k=1,2,3$, in the BCRS, as given in \cite{Dhu2005}. Those BCRS coordinates $%
\left( x_{k},y_{k},z_{k}\right) $, for arbitrary initial conditions, can be
rewritten in terms of rotated keplerian ellipses $\left( x_{ell\ k},y_{ell\
k},z_{ell\ k}\right) $ as 
\begin{eqnarray}
\left( 
\begin{array}{l}
x_{k} \\ 
y_{k} \\ 
z_{k}
\end{array}
\right) &=&\Re ^{-1}\left( 
\begin{array}{l}
x_{ell\ k} \\ 
y_{ell\ k} \\ 
z_{ell\ k}
\end{array}
\right)  \label{BCRS_orbit_coordinate_equations} \\
&&  \nonumber
\end{eqnarray}
$\quad $with\ \medskip \newline
$\Re ^{-1}\equiv \left( 
\begin{array}{lll}
+\cos \Omega _{k}\cos \omega -\sin \Omega _{k}\sin \omega \cos i & -\cos
\Omega _{k}\sin \omega -\sin \Omega _{k}\cos \omega \cos i & +\sin \Omega
_{k}\sin i \\ 
+\sin \Omega _{k}\cos \omega +\cos \Omega _{k}\sin \omega \cos i & -\sin
\Omega _{k}\sin \omega +\cos \Omega _{k}\cos \omega \cos i & -\cos \Omega
_{k}\sin i \\ 
+\sin \omega \sin i & +\cos \omega \sin i & +\cos i
\end{array}
\right) \smallskip $\newline
and $\left( 
\begin{array}{l}
x_{ell\ k} \\ 
y_{ell\ k} \\ 
z_{ell\ k}
\end{array}
\right) \equiv \left( 
\begin{array}{l}
a\left( \cos \Psi _{k}-e\right) \\ 
a\sqrt{1-e^{2}}\sin \Psi _{k} \\ 
0
\end{array}
\right) $\medskip \newline
where \medskip \newline
$
\begin{array}[t]{l}
a=1\ \text{A.U}, \\ 
e=\sqrt{1+\frac{4}{\sqrt{3}}\frac{L}{2a}\cos \nu +\frac{4}{3}\left( \frac{L}{%
2a}\right) ^{2}}-1, \\ 
i=arctg(\frac{\frac{L}{2a}\sin \nu }{\sqrt{3/2}+\frac{L}{2a}\cos \nu })
\end{array}
$\medskip \newline
and $\omega $ are the common semi-major axis, eccentricity, inclination and
argument of the periaster of the three spacecraft orbits, respectively. The
optimal inclination of the LISA\ triangle on the ecliptic is $\nu =\frac{\pi 
}{3}+\frac{5}{8}\frac{L}{2a}$ \cite{Na2006} with $L=5\cdot 10^{9}$m, the
average interferometric arm-length. The longitude of the ascending node, $%
\Omega _{k}$, is particular to a given spacecraft $k$ and is given in terms
of that of the first one with a phase shift $\theta _{k}$: 
\[
\Omega _{k}=\Omega _{1}-\theta _{k}\quad \text{with }\theta _{k}\equiv
-2\left( k-1\right) \frac{\pi }{3}. 
\]
The time parametrization of the orbits is given by the equation of the
eccentric anomaly $\Psi _{k}$ of each spacecraft, 
\begin{equation}
\Psi _{k}-e\sin \Psi _{k}=M_{k}\text{ ,}  \label{eccentric_anomaly_equation}
\end{equation}
with the mean anomaly 
\[
M_{k}=\frac{2\pi }{T}\left( t-t_{0}\right) +M_{k0} 
\]
in terms of the orbital period, $T$ (provided by Kepler's third law, $2\pi
/T=\sqrt{GM/a^{3}}$), and the mean anomaly of spacecraft $k$ at initial time 
$t_{0}$, that is $M_{k0}\equiv M_{k}(t=t_{0})$.\newline
Mean anomalies are related to that of the first spacecraft through the phase
shift: 
\[
M_{k}=M_{1}+\theta _{k}\text{ .} 
\]

The common spacecraft orbit eccentricity being small, the eccentric anomaly
equation (\ref{eccentric_anomaly_equation}) can be developed at 1st order
(with respect to $e$): 
\begin{eqnarray}
\Psi _{k} &\simeq &+\frac{2\pi }{T}(t-t_{0})+M_{10}+\theta _{k}+e\sin \left( 
\frac{2\pi }{T}(t-t_{0})+M_{10}+\theta _{k}\right) \text{ .}
\label{1st_order_eccentric_anomaly_equation} \\
&&  \nonumber
\end{eqnarray}

BCRS position and eccentric anomaly equations used in \cite{Chau2005}
correspond to particular initial conditions $\left( t_{0}=0,\ \omega =3\pi
/2,\ \Omega _{1}=3\pi /2,\ M_{10}=0\right) $ without any planets (which
means that both $t_{0}$ and $M_{10}$ are completely arbitrary). When planets
are modeled, the LISA guiding center, that is the projection of LISA's
center of mass on the ecliptic plane, is supposed to be about 20 degrees
behind the Earth at initial time $t_{0}$. This means that $t_{0}$ and $%
M_{10} $ are no more arbitrary. For example, if the initial mission time is $%
t_{0}=0 $ on epoch Jan 01 2012 at 00:00:00, we find $M_{10}\simeq -.955002$
from keplerian Earth ephemeris for $\omega =3\pi /2,\ \Omega _{1}=3\pi /2$.


\subsection{Relativistic time transformation}

\label{time_transfo}
\qquad We now wish to compute the time transformation between proper time, $%
\stackrel{k}{\tau }$, of the clock on board spacecraft $k$, as a function of
coordinate (BCRS) time $t$, called TCB; since coordinate time is the common
``language'' between the different spacecrafts ($k=1,2,3$) and the time used
in the TDI method.

\begin{figure}[t]
\begin{center}
\includegraphics[width=0.8\textwidth]{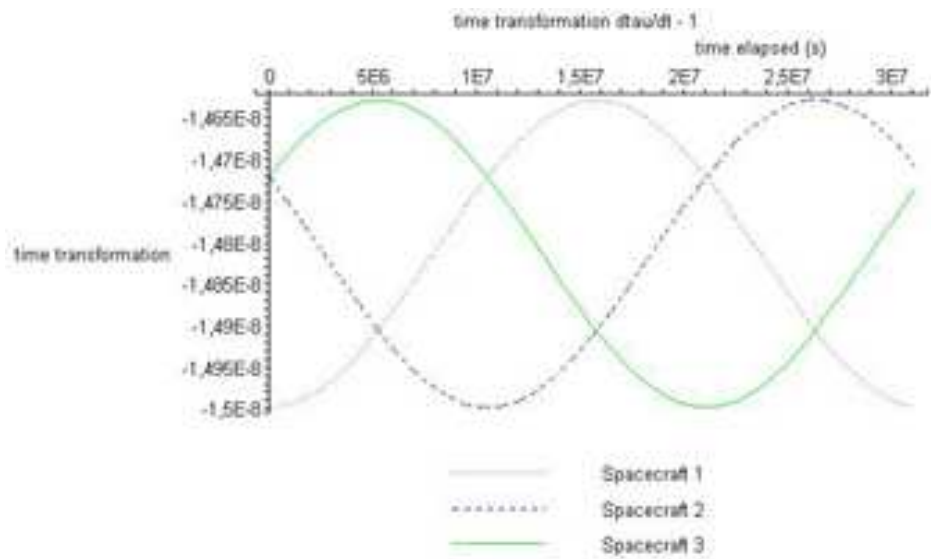}
\end{center}
\caption{Differential proper versus TCB time transformation for spacecraft $k
$ (Eq. \ref{dtau_over_dt}). }
\label{fig_dtau_over_dt_moins_1}
\end{figure}

\begin{figure}[t]
\begin{center}
\includegraphics[width=0.8\textwidth]{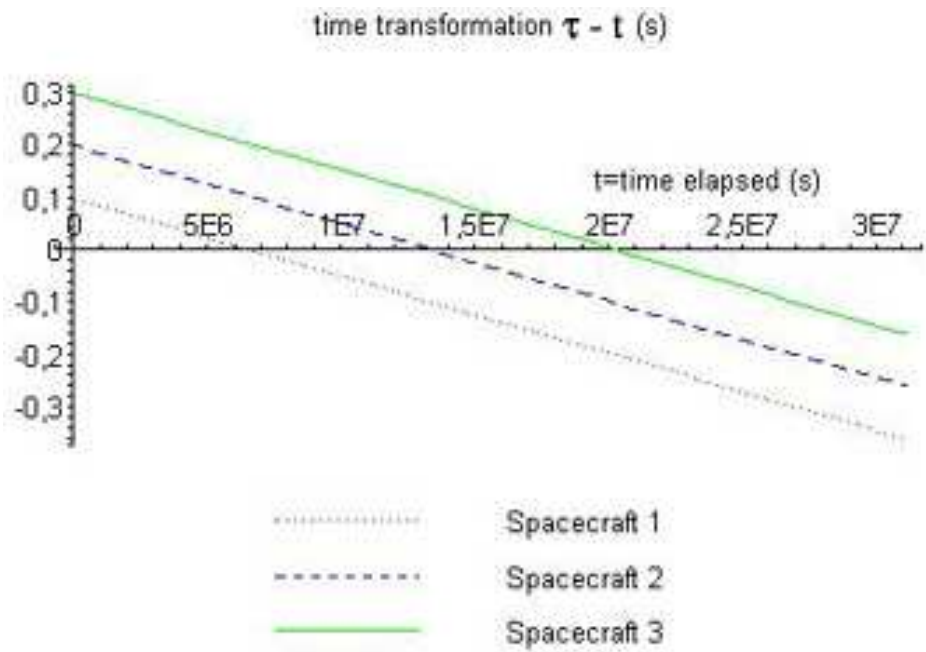}
\end{center}
\caption{Proper versus TCB time transformation (Eq. \ref{tau_moins_t}) for
spacecraft $k=1,$ $2,$ $3$, integrated over a one-year mission. In this
figure, for the sake of the illustration, there is an initial shift of $0.1$%
; $0.2$; $0.3$ s for the clock aboard spacecraft $k=1,$ $2$ or $3$
respectively.}
\label{fig_tau_moins_t}
\end{figure}

\begin{figure}[t]
\begin{center}
\includegraphics[width=0.8\textwidth]{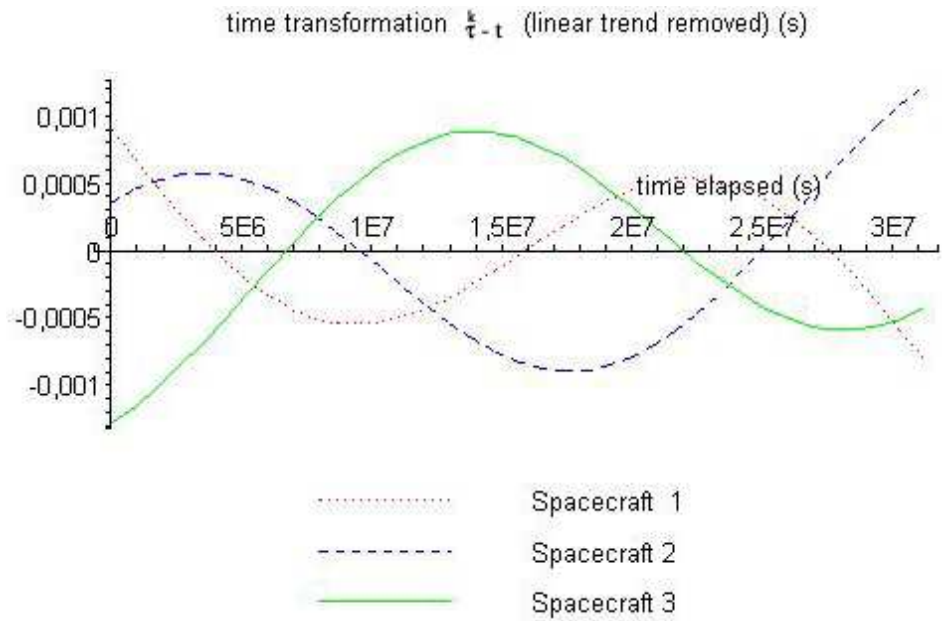}
\end{center}
\caption{Differential proper versus TCB time transformation (Eq. \ref
{tau_moins_t}) for spacecraft $k=1,$ $2,$ $3$, the linear trend is removed
using a least-square fit method.}
\label{fig_tau_moins_t_oscillatory}
\end{figure}

If we consider only the gravitational field due to the Sun, the time
transformation is given by (Figure \ref{fig_dtau_over_dt_moins_1}) 
\begin{eqnarray}
ds^{2} &=&c^{2}d\stackrel{k}{\tau }^{2}\simeq \left( 1-2\frac{w_{k}}{c^{2}}-%
\frac{v_{k}^{2}}{c^{2}}\right) c^{2}dt^{2}  \nonumber \\
&\Rightarrow &d\stackrel{k}{\tau }\simeq \left( 1-\frac{w_{k}}{c^{2}}-\frac{%
v_{k}^{2}}{2c^{2}}\right) dt  \label{dtau_over_dt}
\end{eqnarray}
where $c$ is the speed of light in vacuum, $w_{k}\equiv \frac{GM}{r_{k}}$
with $G$, Newton's gravitational constant, $M$, the mass of the Sun, $r_{k}$
the radial distance relative to the Sun at time $t$ and $v_{k}$ the velocity
of spacecraft $k$ at time $t$ in the Barycentric Coordinate Reference System.%
\newline
We can compute the norm of the satellite keplerian velocity and keplerian
radial distance using equations (\ref{BCRS_orbit_coordinate_equations}),
leading to 
\begin{eqnarray}
v_{k}^{2} &=&\left( \frac{2\pi }{T}\right) ^{2}a^{2}\frac{1+e\cos \Psi _{k}}{%
1-e\cos \Psi _{k}}\ .  \label{velocity_norm} \\
r_{k} &=&a\ \left( 1-e\cos \Psi _{k}\right) \ .  \label{radial_distance_norm}
\end{eqnarray}
\newline
Substituting those expressions in the time transformation (\ref{dtau_over_dt}%
), we obtain 
\begin{eqnarray}
\stackrel{k}{\Delta} &\equiv &\stackrel{k}{\tau }-t\simeq -\frac{GM}{c^{2}a}%
\int \frac{1}{1-e\cos \Psi _{k}}dt-\frac{a^{2}}{2c^{2}}\left( \frac{2\pi }{T}%
\right) ^{2}\int \frac{1+e\cos \Psi _{k}}{1-e\cos \Psi _{k}}dt  \nonumber \\
&\simeq &-\frac{GM}{c^{2}a}\frac{T}{2\pi }\Psi _{k}-\frac{a^{2}}{2c^{2}}%
\frac{2\pi }{T}\Psi _{k}-\frac{a^{2}}{2c^{2}}\frac{2\pi }{T}e\sin \Psi
_{k}+cst  \nonumber \\
&\simeq &\stackrel{k}{\tau _{0}}-t_{0}-\frac{\sqrt{GMa}}{2c^{2}}\left[
3\left( \Psi _{k}-\Psi _{k0}\right) +e\ \left( \sin \Psi _{k}-\sin \Psi
_{k0}\right) \right] \ .  \label{tau_moins_t}
\end{eqnarray}
\newline
The integration constant is given by $\stackrel{k}{\tau }=\stackrel{k}{\tau }%
_{0}$ and $\Psi _{k}=\Psi _{k}(t_{0})\equiv \Psi _{k0}$ at initial
coordinate time $t=t_{0}$. The integration was performed through the change
in variable $dt=T\left( 1-e\cos \Psi _{k}\right) /2\pi \ d\Psi _{k}$, using
the time derivative of the exact expression for the eccentric anomaly (\ref
{eccentric_anomaly_equation}), and Kepler's third law.\newline
From the implicit equation (\ref{eccentric_anomaly_equation}) providing $%
\Psi _{k}(t)$, one can compute $\stackrel{k}{\Delta}$ as a function of time $%
t$ (Figures \ref{fig_tau_moins_t} and \ref{fig_tau_moins_t_oscillatory}).
Alternatively, one could use the 1st order expression (\ref
{1st_order_eccentric_anomaly_equation}).

To obtain the figures in section \ref{figures}, the initial time was chosen
as $t_{0}=0$, the initial offsets of the clocks where $\stackrel{1}{\tau }%
_{0}=0.1$ s$,$ $\stackrel{2}{\tau }_{0}=0.2$ s, $\stackrel{3}{\tau }_{0}=0.3$
s with initial conditions $t_{0}=0,\ \omega =3\pi /2,\ \Omega _{1}=3\pi /2$
and$\ M_{10}=0$. The cumulative $\stackrel{k}{\tau }-t$ effect reaches about
half a second over one year. \newline
In Figure \ref{fig_tau_moins_t_oscillatory}, the linear trend is removed.
The amplitude of the $\stackrel{k}{\Delta}$ oscillatory behavior is $\sim
10^{-3}$ s.


\section{LISA model and TDI technique: a coherent general relativistic approach}

\begin{figure}[t]
\begin{center}
\includegraphics[width=0.9\textwidth]{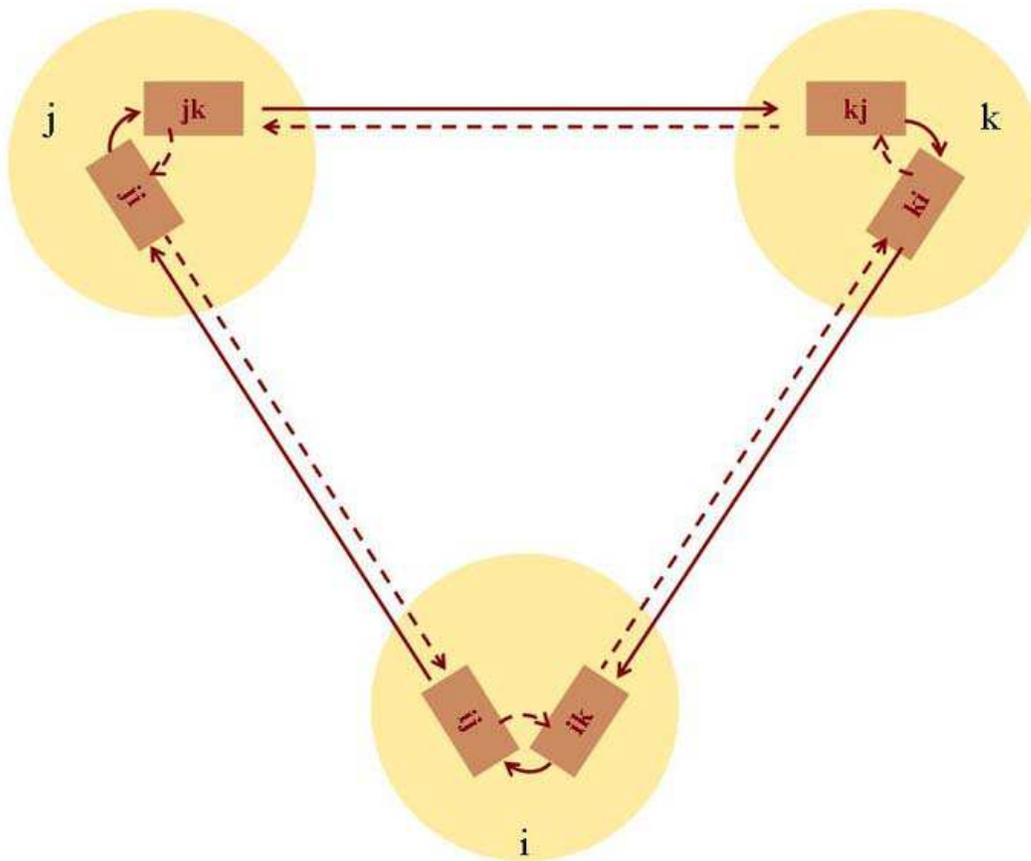}
\end{center}
\caption{Model for the LISA mission: double laser links between spacekraft $%
i,j,k$ and double fiber links between the two lasers aboard each spacecraft.}
\label{fig_LISA_fiber_laser_links}
\end{figure}

We now recall how this proper time of spacecraft $k$ versus TCB
transformation, $\stackrel{k}{\Delta}$, fits in the TDI picture. Let us
consider data flow variables $ij$, collected in a vector, 
\begin{equation}
\overrightarrow{\varepsilon }\equiv (\varepsilon _{31},\varepsilon
_{12},\varepsilon _{23};\varepsilon _{21},\varepsilon _{32},\varepsilon
_{13};\varepsilon _{11},\varepsilon _{22},\varepsilon _{33})\text{\quad } .
\label{data_flow_vector}
\end{equation}
We briefly describe the content of the above vector\footnote{
Our notations $(\varepsilon _{31},\varepsilon _{12},\varepsilon
_{23};\varepsilon _{21},\varepsilon _{32},\varepsilon _{13};\varepsilon
_{11},\varepsilon _{22},\varepsilon_{33})$ correspond to 
$(U_{1},U_{2},U_{3};-V_{1},-V_{2},-V_{3};$ $Z_{1},Z_{2},Z_{3})$ in reference 
\cite{Na2004}} 
using Figure \ref{fig_LISA_fiber_laser_links}. \\
Each fiber data-flow variable, 
\begin{equation}
\varepsilon _{jj}(t)\equiv \left( \Gamma^{ijk}(t)-\Gamma^{kji}(t) \right) /2 
\text{\quad } \text{with }j=1,2,3,  \label{fiber-link_vector}
\end{equation}
consists in an antisymmetric combination of the two data flows at station $j$, 
\begin{equation}
\Gamma^{ijk}(t)\equiv \left. \Gamma^{ijk}\right| _{t\text{, at $j$ with
LF,OB,F,PM noises}} \text{\quad }  \label{fiber-link_frequency_shifts}
\end{equation}
and $\Gamma^{kji}(t)$, that compare the frequencies of the two local lasers (%
$jk$ and $ji$) via fiber links.\\
Each laser-link data-flow variable, 
\begin{equation}
\varepsilon _{ij}(t)\equiv E_{ij}\Lambda^{ki}(t)-\Lambda^{ij}(t) \text{\quad 
} \text{with }i\neq j, j\neq k, k\neq i\text{ and }i\text{, }j\text{, }%
k=1,2,3,  \label{laser-link_vector}
\end{equation}
consists in comparing, at station $j$, the frequency shift between the
incoming laser beam $i\rightarrow j$ and the local laser aligned in
direction $j\rightarrow i$ at station $j,$ 
\begin{equation}
\Lambda^{ij}(t)\equiv \left. \Lambda^{ij}\right| _{t\text{, at $j$ with
LF,Q,OB,PM noises + GW signal + G}} \text{\quad },
\label{laser-link_frequency_shifts}
\end{equation}
with that at station $i$, $\Lambda^{ki}(t)$, considering that the light has
travelled a path $i\rightarrow j$ which translates into a corresponding time
delay $t_{ij}$. In terms of operators, it means dealing with the time-delay
operator $E_{ij}$ so that for a given function of time $f(t)$, 
\begin{equation}
E_{ij}\left( f(t)\right) \equiv f(t-t_{ij})\ .  \label{time_delay_operator}
\end{equation}
The (coordinate-)time transfer between spacecraft $i$ and spacecraft $j$, 
\begin{equation}
t_{ij}\equiv t_{rec\ j}-t_{em\ i}\ ,  \label{time_delay}
\end{equation}
is the barycentric coordinate flight-time of photons between spacecraft $i$
(photon emitted at time $t_{em\ i}$ by spacecraft $i$) and spacecraft $j$
(photon received at time $t_{rec\ j}$ by spacecraft $j$). This quantity is
useful in TDI techniques: data-flow variables are combined with appropriate
time delays to form TDI observables which are (almost) free of LF and OB noises: 
\begin{equation}
TDI\text{ }observable_{l}=\sum_{m=1}^{m_{max}}d_{l}^{m}\varepsilon ^{m}
\label{TDI_observable}
\end{equation}
where 
\begin{equation}
\overrightarrow{d}_{l}=m_{max}-uple\ polynomial\ of\ p\ variables\ E_{ij}(t)
\label{TDI_generator}
\end{equation}
is the $l$-th TDI generator of a chosen generating set with $l=1..l_{max}$.
The variable  $l_{max}$ is the total number of generators in the chosen generating set and 
$m_{max}$ is the number of data-flow variables considered in the vector 
$\overrightarrow{\varepsilon }$. The size, $n$ with $n\leq l_{max}$, of the
smallest generating set of polynomials that cancel LF and OB
noises, as well as $m_{max}$ or $p$ depend on the level of modeling of the
LISA mission.\newline

Indeed, one distinguishes between 1st, 1.5th and 2nd generation TDI.\newline
The First Generation TDI assumes constant (in time) and symmetric time
delays ($t_{ij}=t_{ji}$). This leads to $p=3$ time-delay operators, a
minimum of $n=4$ TDI generators and $m_{max}=6$ data-flow variables (or $9$
if LISA's internal motions are considered). The set of
1st-generation-TDI observables forms the first module of syzygies over a
ring of $p=3$ variables \cite{Dhu2002}. For ideal, perfect, identical (stable,
accurate, without shift nor drift) clocks beating the time $t$ aboard the
three stations, if the time delays $t_{ij}$ are known exactly, within the
1st-generation-TDI assumptions, the LF (and OB) noises are exactly cancelled
in 1st-generation-TDI observables formed from the data-flow variables
recorded at each spacecraft.

The 1.5th Generation TDI still assumes constant time delays, but they are
not reciprocal anymore ($t_{ij}\neq t_{ji}$). Hence, the number of
time-delay operators is doubled $p=6$, while $n=6$ and $m_{max}=9$. The
1.5th-generation-TDI observables still form a module over a ring of $p=6$
variables \cite{Na2004}. Hence, within the 1.5th-Generation-TDI assumptions,
for ideal perfect identical clocks beating the time $t$ and exact knowledge
of $t_{ij}$, the LF and OB noises are still exactly cancelled in
1.5th-generation-TDI observables.

The 2nd Generation TDI relaxes the constant time-delay assumption 
($t_{ij}\equiv t_{ij}(t)$) with respect to the 1.5th Generation TDI.
Consequently, the time-delay operators ($p=6$) do not commute anymore. A new
module still must be found. In reference \cite{Sha2003}, the authors propose
6 new, more complex, generalized Michelson- and Sagnac-type
2nd-generation-TDI observables. For those, the order in which the delays are
applied matters. Those do not remove \emph{exactly} the LF and OB noises, 
even for ideal perfect identical clocks beating the time $t$ and exact knowledge of $t_{ij}$, 
but those bring them down to an acceptable level for LISA.\newline

Let us first consider the different TDI generations with respect to LISA geometry 
(orbitography and laser links) modeling.\\ 
The first-generation-TDI assumptions are only met by a rigid, motion-less (this
implies that no gravitational bodies are around to cause any motion) LISA
constellation model. At this level of modeling, each time delay is given by
the corresponding constant interferometric arm-length, slightly different
from the generic arm-length $L$: $t_{ij}=L_{ij}/c$ with $L_{ij}\equiv 
\overrightarrow{x}_{j}-\overrightarrow{x}_{i}$.

The 1.5-generation-TDI assumptions allow for a rotating (around its center
of mass and around the barycenter) rigid LISA. At that stage, the Sagnac 
\cite{Na2005} and aberration effects cause the non reciprocity of
time-delays \cite{Co2003}. Indeed, for a same path-length, light rays travelling clockwise
and counterclockwise do not take the same time (Sagnac effect).
Additionally, there is a motion of the arm-length $ij$ with respect to the
barycenter causing an aberration effect. The model for LISA corresponding to
1.5-generation-TDI assumptions consists in \emph{classical} motion
(keplerian) of the three stations in the gravitational field of the Sun up
to first order in eccentricity ($e_{LISA}\cong 0.0096$).

The true LISA requires 2nd Generation TDI. A \emph{classical} keplerian
orbital motion around the Sun, if higher orders in eccentricity are
considered, causes a flexing of the constellation: $L_{ij}(t)$ is a simple
periodic function of $t$ if only the Sun is considered, the so-called breathing of
the triangle. If the newtonian perturbation of planets is considered, the
flexing becomes more complex.\newline
Moreover, if a gravitational \emph{relativistic} description of photon
time transfer is adopted, $t_{ij}$ has additional time-dependant contributions
due to gravitational relativistic effects. Indeed, in reference 
\cite[Section III]{Chau2005}, a native, coherent, gravitational relativistic
description of the laser link provides 
$t_{ij}=\stackrel{(0)}{t}_{ij}+\stackrel{(1/2)}{t}_{ij}+\stackrel{(1)}{t}_{ij}$, 
as a function of the positions and velocities of the emitting and receiving 
spacecraft at emission time. 
The 0th order in $GM/c^2\sim v^2/c^2$, $\stackrel{(0)}{t}_{ij}$, is the 
classical time taken by light to travel $L_{ij}(t)$ at velocity $c$, 
the 1/2th order contains the Sagnac and aberration
effects and the 1st order, light deflection or the so-called Shapiro delay.
In reference \cite[ Section VI]{Chau2005}, these $\stackrel{(0)}{t}_{ij}$, 
$\stackrel{(1/2)}{t}$ and $\stackrel{(1)}{t}_{ij}$ contributions to the photon
flight time were evaluated numerically to $5\cdot 10^{9}$ m/c $\approx 16.7$ s
plus a flexing of amplitude of $48000$ km/c $\approx 0.17$ s, $\sim3\cdot 10^{-3}$ s
and $< 10^{-7}$ s respectively, assuming \emph{classical} ephemerides of the
three LISA stations. \newline
In reference \cite{Pi2007}, \emph{relativistic} ephemerides of the three LISA
stations are provided. The authors compute numerically $L_{ij\text{
relativistic}}(t)-L_{ij\text{ classical}}(t)$ and show that it reaches up to
about $3$ km over a year, corresponding to an extra $\sim10^{-5}$ s in 
$\stackrel{(0)}{t}_{ij}$.\newline
A realistic model of LISA's orbital motion also leads to geometric frequency
shifts in expression (\ref{laser-link_frequency_shifts}). In \cite[ Section V]{Chau2005},
the native, coherent, gravitational relativistic description of the laser
link also provides 
$\Lambda^{ij}_{ G}=\stackrel{(1/2)}{\Lambda}^{ij}_{ G}+%
\stackrel{(1)}{\Lambda}^{ij}_{ G}+\stackrel{(3/2)}{\Lambda}^{ij}_{ G}$, 
as a function of the positions and velocities of the emitting and receiving 
spacecraft at emission time. 
The 1/2th order is the Doppler shift due to the
relative velocities of the emitting and receiving stations in the special
relativistic framework; it is reciprocal. Higher order terms that contain
gravitational relativistic effects such as the Einstein gravitational
Doppler shift in $\stackrel{(1)}{\Lambda}^{ij}_{ G}$, are non reciprocal.
In reference \cite[ Section VI]{Chau2005}, these contributions to photon 
frequency shift were evaluated numerically assuming
\emph{classical} ephemerides of the three LISA stations, leading to 
$\stackrel{(1/2)}{\Lambda}^{ij}_{ G}\sim7\cdot10^{-8}$, 
$\stackrel{(1)}{\Lambda}^{ij}_{ G}\sim2\cdot10^{-13}$ and 
$\stackrel{(3/2)}{\Lambda}^{ij}_{ G}\sim2\cdot10^{-14}$. 
Hence, the geometric frequency shift is irrelevant to LISA, owing to the
mission's detection frequency bandwidth.\newline

\begin{figure}[b]
\begin{center}
\includegraphics[width=0.9\textwidth]{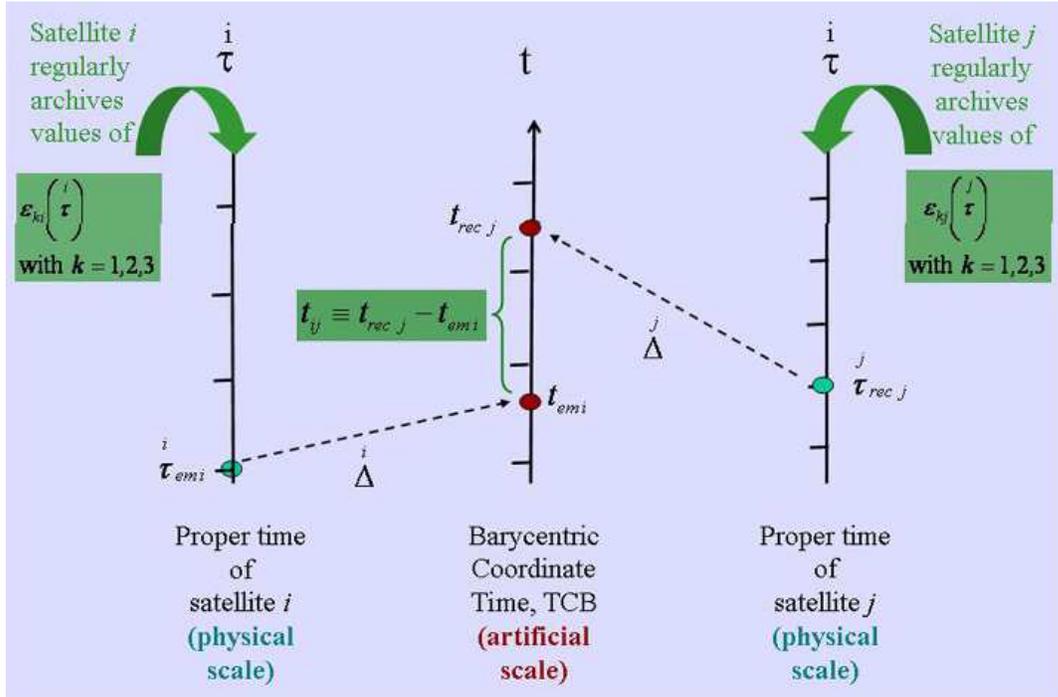}
\end{center}
\caption{The TDI method is developed in terms of the coordinate time scale, 
$t$, associated with the Barycentric Coordinate Reference System which is an
artificial time scale (it can only be computed, not measured). Since each
spacecraft $i$ or $j$ has its own physical time scale in terms of the proper
time beaten by its clock $\stackrel{i}{\tau }$ or $\stackrel{j}{\tau } $, a
coordinate time $t$ is the natural ``common language''. However, events
(such as emission or reception of a signal) and proper frequency shifts are
recorded locally at each spacecraft on their own proper time scale. Hence, a
time transformation, $\stackrel{i}{\Delta}$ or $\stackrel{j}{\Delta}$, is
required with $\stackrel{k}{\Delta} \equiv \stackrel{k}{\tau }-t$.}
\label{relativistic_timescales}
\end{figure}

Let us now consider the TDI analysis from the point of view of LISA time scales. 
TDI-observables (1st, 1.5th and future module
for 2nd generation) should cancel exactly the LF and OB noises for ideal,
perfect, identical clocks beating the time $t$ aboard the three stations.
However, general relativity taught us that the physical times scales are
the proper times scales $\stackrel{1}{\tau }$, $\stackrel{2}{\tau }$ and 
$\stackrel{3}{\tau }$; that is the time beaten by the clocks aboard
spacecraft $1$, $2$ and $3$ respectively, \emph{if} those are \emph{not}
constrained/synchronized\footnote{%
Indeed, equation (\ref{tau_moins_t}) for stations $k=1..3$ orbiting the Sun
is similar to the GPS clock synchronization \cite{Ash2003} applied to clocks
aboard the GPS constellation orbiting the Earth.}. 
Strictly speaking, $\stackrel{1}{\tau }$, $\stackrel{2}{\tau }$ and 
$\stackrel{3}{\tau }$ are different time scales which are not directly compatible, 
like apples and oranges. The natural common language between those time scales is a time
coordinate $t$ such as the one associated with the Barycentric Coordinate
Reference System, called TCB. In the LISA mission, the events, which should
thus be recorded in the corresponding proper time scale aboard a given LISA
spacecraft $k$, are the different emission or reception event at that
spacecraft. The locally measured frequency shifts (against the local proper
frequency) should also be regularly recorded on board 
according to the local proper time scale (see Figure \ref
{relativistic_timescales}). Each spacecraft $k$ data-record should thus contain 
$\stackrel{k}{\tau }_{n\ em\ k}$, $\stackrel{k}{\tau }_{n\ rec\ k}$, 
\begin{equation}
\Gamma^{jki}(\stackrel{k}{\tau })\equiv\left. \Gamma^{jki}\right| _{%
\stackrel{k}{\tau }\text{, at $k$ with LF,OB,F,PM noises}}\quad \text{,}
\label{Archived_fiber-link_frequency_shifts}
\end{equation}
and 
\begin{equation}
\Lambda^{jk}(\stackrel{k}{\tau })\equiv\left. \Lambda^{jk}\right| _{%
\stackrel{k}{\tau }\text{, at $k$ with LF,Q,OB,PM noises + GW signal + G}%
}\quad \text{,}  \label{Archived_laser-link_frequency_shifts}
\end{equation}
where we adopted the notation $\stackrel{k}{\tau }_{em/rec\ k}$ for the n-th
emission/reception event at spacecraft $k$ on the time scale $\stackrel{k}{\tau }$. 
Consequently, to apply the TDI analysis, i.e. to compute (\ref
{TDI_observable}) from (\ref{fiber-link_vector}, \ref
{fiber-link_frequency_shifts}, \ref{laser-link_vector}, \ref
{laser-link_frequency_shifts}), we need to use equation (\ref{tau_moins_t}) 
to convert from time scale $\stackrel{k}{\tau }$, in which the
observed/simulated frequency shifts (\ref
{Archived_fiber-link_frequency_shifts}, \ref
{Archived_laser-link_frequency_shifts}) should be recorded, to time scale $t$.\\

To summarize this section: a coherent general relativistic approach of TDI analysis is necessary. 
This means general relativistic modeling of laser links \cite{Chau2005} (that is $t_{ij}$, 
since geometric frequency shifts can be omitted owing to LISA frequency bandwidth), 
LISA station orbitography \cite{Pi2007} and time scales, which is the subject of the present paper.\medskip


\section{LISA Code: a LISA Simulator}


\qquad A LISA simulator aims at an intimate understanding of the mission, providing an overview of data analysis. 
This means that the simulator will be a precious tool to find and test strategies to detect GW, to test the specificities of interferometric configurations for that purpose, to test the response of the detector (see for example LISA SIMULATOR \cite{Co2007}, SYNTHETIC LISA \cite{Va2005} or LISA Code \cite{Pe2006}). \\

However, before getting to detection strategies, we have seen that the data must go through the TDI pre-processing analysis in order to remove (most of) LF and OB noises. A sine-qua-non condition for the LISA mission to work is for the TDI method to be efficient. \\
Limitations on the effectiveness of the TDI technique come from the fact that the considered TDI generation is more or less adequate to the real orbitography and to the real laser links in the LISA mission. They also come from secondary noises (Q, F, PM, USO -Ultra-Stable Oscillator, that is the clock aboard LISA spacecraft-...) affecting the measurement, and certainly from the finite accuracy/precision of the quantities required to construct the corresponding TDI observables (see \cite[Section V]{Tin2003}).

Consequently, one aim of LISA Code is also to test TDI pre-processing efficiency for a realistic model of the LISA mission (orbitography, laser links, time scales) that is a coherent general relativistic rather than classical model, in presence realistic noises: LF, OB, Q, F, PM, USO. 
With that respect, the question that LISA Code addresses is: Is the residual LF plus OB noise left in a given TDI-generation observable acceptable?
Presently, LISA Code uses a classical orbitography model (including constellation breathing plus rotation around its center of mass and around the barycenter), such as described in \cite{Dhu2005} and allows also for a static LISA or a rigidly rotating LISA. LISA Code adopts a general relativistic laser links description according to \cite{Chau2005} to compute $t_{ij}$. LISA Code also contains LF, OB, PM and USO (shift and drift) noises; standard TDI observables of first and second TDI generation, as well as a procedure to compute arbitrary TDI observables. 
Each of the three USO noises \emph{initially} represents \emph{intrinsic} offset, difference in rate and drift of the clocks aboard one of the three LISA spacecraft. But via this USO-noise input in LISA Code, the proper time versus TCB time transformations, $\stackrel{k}{\Delta}$, described in Section \ref{time_transfo}, can also be introduced. There will then be interesting questions to answer: 
How will the precision of LISA ephemerides, used in LISA Code to compute both $t_{ij}$ and $\stackrel{k}{\Delta}$, affect the efficiency of a given TDI generation? And how will the constraints on LISA ephemerides accuracy derived with LISA Code compare with the theoretical constraints provided by Tinto et al. (about $16$ m for arm-length \cite[Section V A, (29) and (33)]{Tin2003} and about $50$ ns for clock synchronization \cite[Section V B, (39) and (40)]{Tin2003}) for the 1.5th-generation standard TDI observables?


\section{Conclusions}


\qquad We provided the relativistic analytical and numerical estimates for
spacecraft proper times versus coordinate (BCRS) time transformation, for a
classical orbitography in the gravitational field of the Sun. We showed that
the difference between station proper time and BCRS time reaches about $0.5$
s over a one-year mission and exhibits an oscillatory trend with a maximal
amplitude of about $0.0014$ s. The difference in rate of station proper time
versus BCRS time is of the order of $1.5\cdot 10^{-8}$. \\
We then recalled
how this time transformation fits in a general relativistic approach of 
Time Delay Interferometry, necessary to lower the laser frequency noise and optical bench noise
down to the gravitational wave detection threshold. 

\newpage

\section{FIGURES}


\label{figures} \vspace{1cm} 

\pagebreak

\clearpage

\ack{The research work presented here was carried out at the Observatoire de
la C\^{o}te d'Azur financed by CNES (Centre National d'Etudes Spatiales,
France) post-doctoral grant. 
The author also wishes to thank Dr J-Y. Vinet and Dr B. Chauvineau 
(Observatoire de la C\^{o}te d'Azur) for fruitful discussions.}\bigskip


\end{document}